\documentclass[runningheads]{llncs}
\usepackage[T1]{fontenc}

\usepackage{caption}
\usepackage{subcaption} 
\usepackage{graphicx}
\usepackage{hyperref}
\usepackage{color}

\newcommand{\MuDoC}[0]{MuDoC}
\newcommand{\TeDoC}[0]{TexDoC}
\newcommand{\INC}[0]{INC}
\newcommand{\ANA}[0]{ANA}

\begin{document}
\title{Towards a Multimodal Document-grounded Conversational AI System for Education}
\titlerunning{Multimodal Document-grounded Conversational AI for Education}
\author{Karan Taneja\inst{1} \and
Anjali Singh\inst{2} \and
Ashok K. Goel\inst{1}}
\authorrunning{Taneja et al.}
\institute{Georgia Institute of Technology, Atlanta GA 30332, USA \and
University of Texas at Austin, Austin TX 78712, USA \\
\email{\{karan.taneja,ashok.goel\}@cc.gatech.edu, singhanj@umich.edu}}

\maketitle              

\begin{abstract}
Multimedia learning using text and images has been shown to improve learning outcomes compared to text-only instruction. But conversational AI systems in education predominantly rely on text-based interactions while multimodal conversations for multimedia learning remain unexplored. Moreover, deploying conversational AI in learning contexts requires grounding in reliable sources and verifiability to create trust. We present \textbf{MuDoC}, a \underline{Mu}ltimodal \underline{Do}cument-grounded \underline{C}onversa-tional AI system based on GPT-4o, that leverages both text and visuals from documents to generate responses interleaved with text and images. Its interface allows verification of AI generated content through seamless navigation to the source. We compare \MuDoC~to a text-only system to explore differences in learner engagement, trust in AI system, and their performance on problem-solving tasks. Our findings indicate that both visuals and verifiability of content enhance learner engagement and foster trust; however, no significant impact in performance was observed. We draw upon theories from cognitive and learning sciences to interpret the findings and derive implications, and outline future directions for the development of multimodal conversational AI systems in education.
\end{abstract}

\section{Introduction}

Multimedia learning involves processing verbal and visual elements and is considered more effective for transfer and retention compared to learning only from verbal content \cite{mayer_multimedia_2002}.
With recent advancements in AI, multimodal dialog systems \cite{feng_mmdialog_2023,kong_tiger_2024} have received increasing attention, but their application to multimedia learning remains under-explored.
Proprietary AI systems 
like ChatGPT 
are used by students for information seeking and learning \cite{kim_examining_2025}
but these systems use text-to-image generation \cite{saharia_photorealistic_2022} for visual content which is largely unreliable for learning contexts, or 
show relevant image-search results from the web.

Our work explores the unique opportunities extended by multimodal docume-nt-grounded AI systems in educational contexts. Our focus is on supporting learners in problem-solving within the context of formative assessments.
To ensure the reliability and trustworthiness of the responses generated by such systems, it is crucial that they provide information from reliable sources, such as textbooks and course slides.
For instance, existing conversational AI systems such as Jill Watson \cite{taneja_jill_2024,kakar_jill_2024} and PET \cite{wolfel_knowledge-based_2024} ground their responses in instructor-approved materials or predefined template-based slides to answer student questions.
The documents used by these systems typically contain visuals to support understanding and engagement, especially in STEM fields. Yet, these visuals have not been utilized by existing systems to generate multimodal responses.
Towards this end, we present \textbf{\MuDoC}, a \underline{Mu}ltimodal \underline{Do}cument-grounded \underline{C}onvers-ational AI system, for multimedia learning. \MuDoC~generates responses interleaved with text and images in a format similar to news or blog articles.
Based on GPT-4o, it uses past chat as context to understand user queries and responds based on retrieved visuals and text from documents. 
\MuDoC's user interface (UI) allows seamless navigation to image and text sources to enable verification of AI-generated content, thereby fostering trust in its responses \cite{thielsch_trust_2018}.
Our work is a step towards creating trustworthy multimodal intelligent documents and employing them in learning environments.
To supplement this paper, we are also releasing detailed technical documentation, a demo video, and sample conversations with \MuDoC~based on the Knowledge-based AI e-book\footnote{Conversation Samples (Google Doc): \href{https://tinyurl.com/MuDoCSamples}{tinyurl.com/MuDoCSamples}\label{footnote:samples-and-documentation}\\Documentation/Demo (OSF): \href{https://tinyurl.com/MuDoC-OSF}{tinyurl.com/MuDoC-OSF}}.

We compare \MuDoC~to a text-only system through a user study with students from a large public university to understand the differences in learner engagement, trust in the AI systems, and their subsequent performance on problem-solving tasks.
Prior studies in multimedia learning research have isolated the learning and the evaluation process as two separate steps \cite{mayer_multimedia_2002}.
However, we consider a more authentic scenario where students solve problems while learning by asking questions to a conversational AI system.
This setting requires students to self-regulate their learning \cite{winne1998studying} as they form questions, read AI responses, and solve problems in a reasonable time frame.
Feedback from the students shows that both visuals and verifiability of AI content support learner engagement and trustworthiness in multimodal conversations,
though results show no significant difference in performance.
Our analysis also provides insights into the cognitive processing involved in this authentic context of multimedia learning, based on which we provide directions for future research.

This paper makes three main contributions: 
(1) We present MuDoC, a multimodal and document-grounded conversational AI system that supports multimedia learning.
(2) We examine \MuDoC's impact on learner engagement, trust, and performance on problem-solving tasks by comparing it to a text-only system.
(3) We present our findings and ground them in theories from cognitive and learning sciences to understand implications for multimodal AI in education.

\section{Related Work}\label{sec:related-work}

\textbf{Multimedia Learning:}
Multimedia learning theory or MLT \cite{mayer_multimedia_2002} posits that building connections between verbal and visual representations of learning content leads to improved transfer and higher cognitive retention.
The cognitive model supporting MLT describes filtering and processing of verbal and visual content in the brain as two independent channels, each with limited capacity. 
These two sources of information are organized and integrated with existing knowledge to build coherent mental representations.
MLT is also closely related to cognitive load theory or CLT \cite{sweller_cognitive_2011} which suggests that we have a limited short-term working memory for processing information.
On the other hand, our long-term memory is unlimited and plays a crucial role in learning as it provides the foundation to working memory from which new knowledge is created and stored.
In our work, we study the impact of multimodal AI on learning to understand if multimedia effect holds true in conversations with AI systems for problem-solving. 
\textit{Goal-free effect} in CLT \cite{sweller_cognitive_2011} suggests that such goal-oriented learning leads to increased cognitive load as learners perform means-end analysis during learning,
and this issue can be intensified further due to time constraints.
We aim to understand the use of conversational AI in this realistic setting where cognitive load is high as students learn while solving problems.

\textbf{Grounded Conversational AI:} 
Jill Watson \cite{taneja_jill_2024,kakar_jill_2024}, based on OpenAI LLMs, is a conversational AI for classrooms.
It uses a retrieval-augmented generation (RAG) pipeline to answer student questions 
by retrieving relevant information and prompting the LLM to generate text-based responses.
Similarly, PET \cite{wolfel_knowledge-based_2024} is a dialog system to use predefined template-based slides for answering student questions using GPT-4.
The chat window is placed within the slide viewer and allows searching relevant slides.
Intelligent textbooks proposed by Olson et al. \cite{olson_textbook_2025} also display a chat window adjacent to documents but do not use visuals or allow navigation to source content.   
In \MuDoC, we use a RAG pipeline and extract figures from documents during the preprocessing step to utilize them for generating interleaved text and image responses.
\MuDoC's interface also allows users to jump to source paragraphs and images in the document.
Other related work has explored text-based document-grounded dialog \cite{feng_doc2dial_2020}, visually-grounded chat \cite{lv_kosmos-25_2024}, and multimodal document understanding \cite{xu_layoutxlm_2021,ye_mplug-docowl_2023}.

\textbf{Multimodal Dialog:}
Many multimodal dialog datasets have been proposed in recent years
including task-oriented \cite{saha_towards_2018}, 
and open-domain datasets \cite{zang_photochat_2021}.
These datasets contain a single or a few relevant images with short utterances describing the content. 
Large scale models for interleaved image and text generation such as DreamLLM \cite{dong_dreamllm_2024}, 
and
MiniGPT-5 \cite{zheng_minigpt-5_2024} 
can generate images and text for visual storytelling and recipe generation  applications.
\MuDoC~is designed to generate \textit{document-grounded} multimodal responses in \textit{conversations}.
While its responses paraphrase original text, it uses images directly extracted from documents for information accuracy and technical correctness.

\section{Multimodal Document-grounded Conversational AI}\label{sec:methods}

We initially introduced \MuDoC~at the AAAI-MAKE  Symposium \cite{taneja_mudoc_2025} where we described details about its preprocessing and retrieval augmented generation pipeline and its user interface.
\MuDoC~processes user queries as a part of conversation to retrieve relevant text and images from documents to generate a multimodal response. 
Its responses contain interleaved images and text in a format similar to news or blog articles. 
The preprocessing and response generation pipeline are outlined in Fig. \ref{fig:preprocessing_and_response_gen}. 
Technical details are provided in the supplementary material\footref{footnote:samples-and-documentation}.

\begin{figure}[t]
    \centering
    \includegraphics[width=\linewidth]{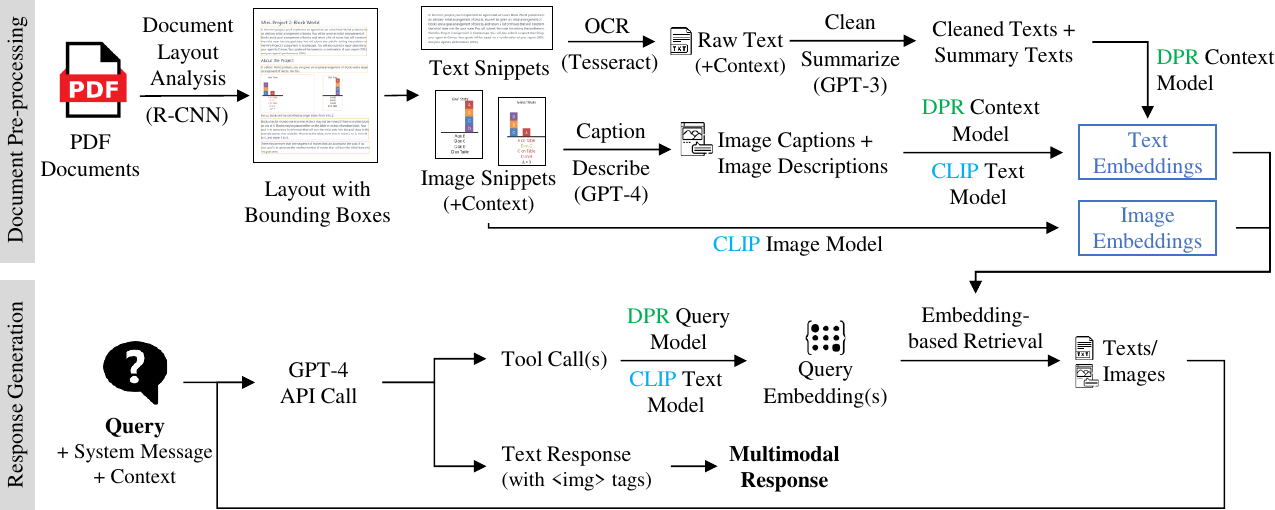}
    \caption{\MuDoC's Document Pre-processing and Response Generation Pipeline. 
    Details and demo available in supplementary material\footref{footnote:samples-and-documentation}.
    }
    \label{fig:preprocessing_and_response_gen}
\end{figure}

\subsection{Document Pre-processing}

The pre-processing step involves building an index of text and images that can be used for embedding-based retrieval during response generation. 
Document Layout Analysis (DLA) is used to identify and classify the regions of interests in a document.
A \href{https://layout-parser.readthedocs.io/en/latest/notes/modelzoo.html}{layout parser} based on Region-based Convolutional Neural Network (R-CNN) \cite{he_mask_2017} trained on PubLayNet dataset \cite{zhong_publaynet_2019} is used to predict bounding boxes and classify them into titles, texts, figures, lists and tables. 
After extracting layout snippets in image format, each snippet is passed through \href{https://tesseract-ocr.github.io/tessdoc/Installation.html}{Tesseract OCR Engine}
to obtain raw text.
The raw text from OCR model can have typos and poor formatting 
which are corrected by prompting GPT-3.5 (turbo-16k-0613) to generate \textit{cleaned text}
as well as a \textit{summary text} which
provides a shorter variation of text for additional text embeddings.
For image snippets, image captions and descriptions are generated 
using GPT-4 (vision-preview) model
to provide more context for referring to images during response generation.

For embeddings, Dense Passage Retrieval (DPR) \cite{karpukhin_dense_2020} models consist of two encoder models viz. query model and context model, which are aligned to output similar embeddings for a query and context if the context can answer the query. 
The text embeddings are pre-computed for retrieval using 
the \href{https://huggingface.co/facebook/dpr-ctx_encoder-multiset-base}{context model} for raw text, cleaned text, summary text, image captions, and image descriptions.
Similar to DPR, Contrastive Language-Image Pre-training (CLIP) \cite{radford_learning_2021} models have two aligned encoder models where one encodes images and the other encodes textual descriptions. 
The \href{https://huggingface.co/openai/clip-vit-base-patch32}{image model} is used to pre-compute image embeddings, and text model is used for caption and description embeddings.

\subsection{Response Generation}

Given a query, the system can perform text and/or image retrieval before generating a response grounded in the source documents.
For text retrieval, the query DPR embedding is computed using 
\href{https://huggingface.co/facebook/dpr-question_encoder-multiset-base}{query model} and cosine similarities with pre-computed context embeddings are used to 
return the top five texts as text retrieval output. 
For image retrieval, both DPR query and CLIP text embeddings are used to find the top five images.

GPT-4o (2024-08-06) model is used to generate responses and its 
\href{https://platform.openai.com/docs/guides/function-calling#structured-outputs}
{tool/function calling feature} is employed for invoking text and image retrieval function calls. 
Note that GPT-4o API response either contains a text response or a \texttt{tool\_call} response with required inputs. 
The text response is returned to user while, 
for \texttt{tool\_call}s, the function call is performed and the results are returned with another API call to GPT-4o.
The model is provided with two search tools viz. \texttt{search\_texts} and \texttt{search\_images}. Both these functions have \texttt{search\_query} as the only input and both return the retrieved content.
The system message instructs GPT-4o to use HTML image tags with image addresses mentioned in the retrieval output. 
When displaying the response to users, HTML image tags are replaces with real images and generated captions are added below each image.

\subsection{User Interface}

\begin{figure}[t]
    \centering
    \begin{subfigure}[b]{\textwidth}
        \centering
        \includegraphics[width=\textwidth]{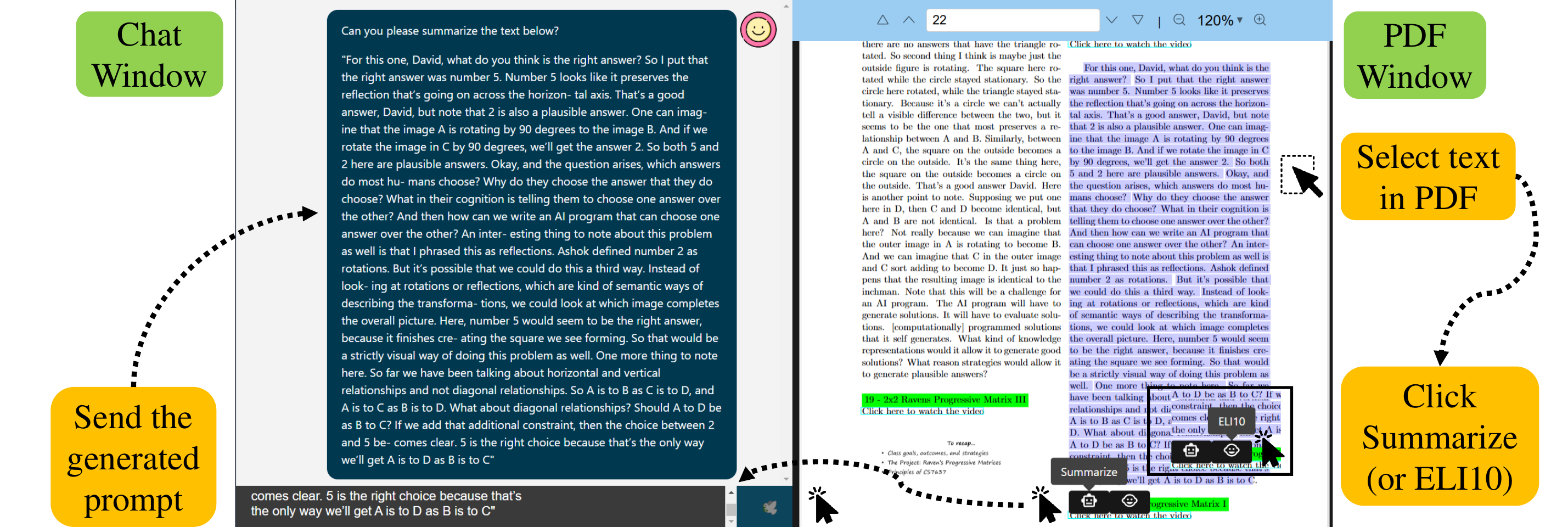} 
        \caption{Chat-PDF Display and Summarize/Explain-it-Like-I'm-10 (ELI10)}
        \label{fig:ui-summarize}
    \end{subfigure}
    \hfill 
    \begin{subfigure}[b]{0.48\textwidth}
        \centering
        \includegraphics[width=\textwidth]{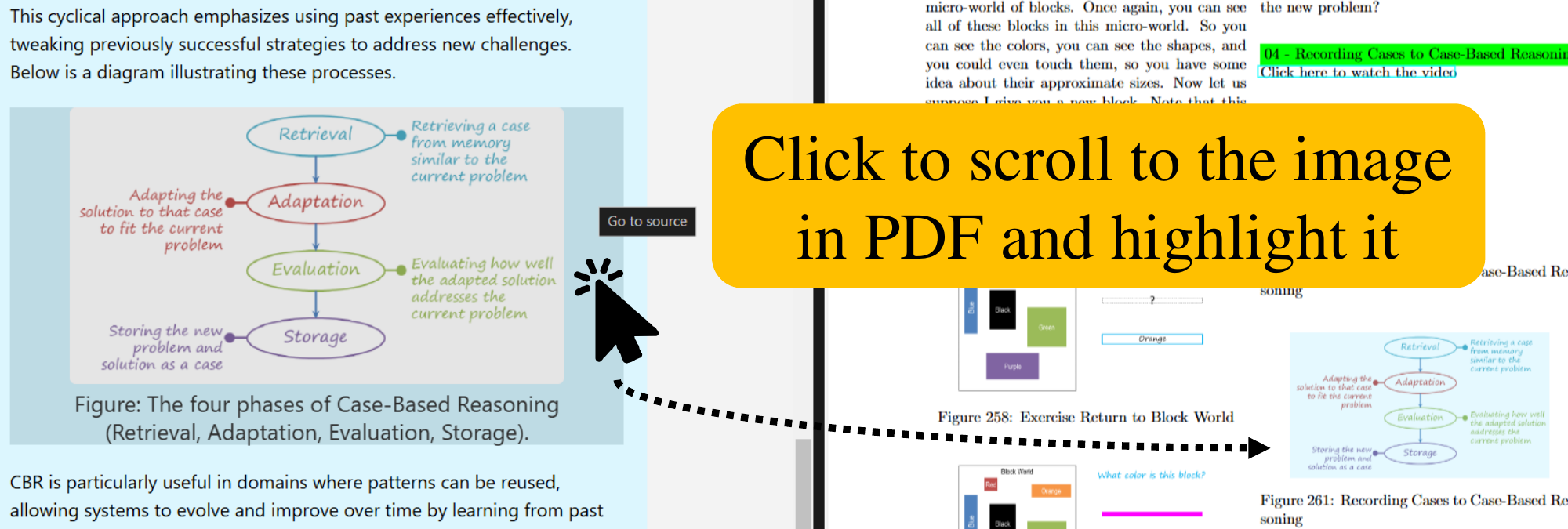} 
        \caption{Clickable Images for Navigation}
        \label{fig:ui-image-highlight}
    \end{subfigure}
    \hfill 
    \begin{subfigure}[b]{0.48\textwidth}
        \centering
        \includegraphics[width=\textwidth]{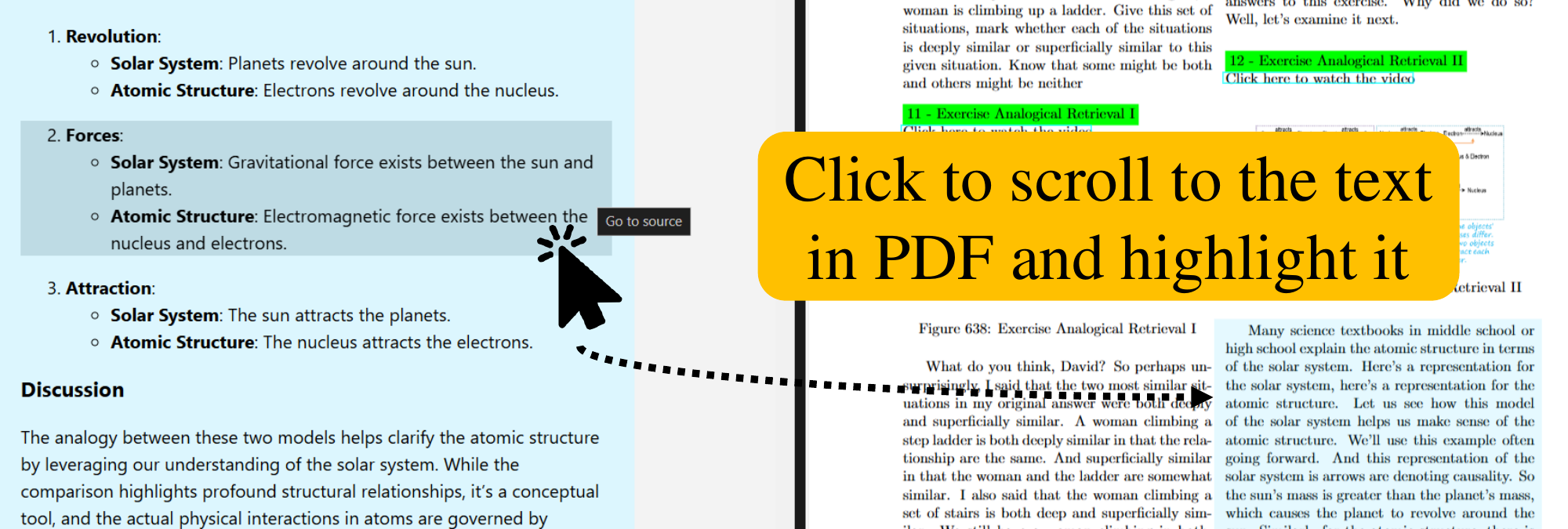} 
        \caption{Clickable Text for Navigation}
        \label{fig:ui-text-highlight}
    \end{subfigure}
    \caption{User Interface Design and Features}
    \label{fig:ui-interface}
\end{figure}

\MuDoC's trust-oriented UI supports interactive features that engage students in exploring long documents.
The UI shows a chat window on the left with an input text-box at the bottom, and
the source document on the right (see Fig. \ref{fig:ui-summarize}).
The front-end is built using \href{https://react.dev/}{ReactJS} framework and served through a \href{https://flask.palletsprojects.com/en/}{Flask} application that also handles response generation.
The token stream from OpenAI API is forwarded to front-end over HTTP using  \href{https://developer.mozilla.org/en-US/docs/Web/API/Server-sent_events/Using_server-sent_events}{Server-Sent Events (SSE)} to reduce the time to first token.
While the back-end is working on retrieval and API calls, UI display messages indicating the system status such as `Gathering information', `Retrieving text/images for [query]', and `Generating a response'. 
\href{https://mozilla.github.io/pdf.js/}{PDF.js} PDF-viewer supported by Mozilla is used to display documents.

\textbf{Summarize and Explain-it-Like-I'm-10 (ELI10):}
A user can select text in the document to view and select `Summarize' and `ELI10' options (Fig. \ref{fig:ui-summarize}).
Each option creates a prompt in the chat text-box where it can be edited before entering it to receive a response.
Summarization leads to a shorter text 
while ELI10 leads to a simpler low-jargon explanation. 
Together, these features support readability and convenience for students in reading long and complex documents.

\textbf{Navigation Using Images and Text:} 
The images and paragraphs in a multimodal response appear clickable through color change and tooltip text (Fig. \ref{fig:ui-image-highlight}, \ref{fig:ui-text-highlight}). 
When these elements are clicked, the document is scrolled to the page where the figure or a similar text appears and the appropriate part of the page is highlighted for three seconds.
For images, the bounding box data stored during pre-processing is used to accomplish this.
For texts, paragraphs are matched to text snippets in the document based on DPR embeddings similarity.
These navigation features help students examine and understand the AI response in context of the document. 
Source attribution, as a form of explanation, also builds trust in AI-generated content as learners can verify AI responses.

\section{Experiment Design}\label{sec:experiments}
To understand the impact of multimodality and interactive features of \textbf{\MuDoC}, 
we designed a within-subjects research study 
where we compare \MuDoC~against a baseline system called 
\textbf{\TeDoC}.
\TeDoC~is the \underline{Tex}t-only variant of \MuDoC~with the same backbone, except
it does not have an image retrieval pipeline or image-related instructions in prompts. 
Further, it only displays a chat window similar to ChatGPT and the PDF document is provided in a separate browser tab. 
The study protocol was approved by our Institute Review Board.

\textbf{Participants:} 
We recruited n=30 graduate (Masters and PhD) students from 
Georgia Institute of Technology (Atlanta, GA, USA)
to participate in the study. 
We screened prospective participants to select those who have finished at least one course in ML or AI domain, but have never enrolled for the Knowledge-based AI course on which the problems for our study were based.

\textbf{Method:}
We invited participants for a 90-minutes in-person session where they solved two subjective analytical problems using pen and paper while asking questions to \TeDoC~and \MuDoC.
The textbook provided to the participants is based on Knowledge-based AI course and contains numerous visuals explaining concepts and algorithms taught in the course. 
Each participant solved the same warm-up task related to \textit{semantic networks} in 7 minutes to familiarize themselves with \TeDoC. 
They were asked to solve the first problem in 20 minutes by asking questions to \TeDoC~system and writing their solutions.  
Then, a transfer problem based on \textit{incremental concept learning} (\textbf{\INC}) or \textit{analogical reasoning} (\textbf{\ANA}) was assigned randomly.
Next, they spent 7 minutes to familiarize themselves with \MuDoC~to solve the same semantic network warm-up problem, but they were additionally instructed to try all \MuDoC~features and ask clarification questions about them.
After this second warm-up task, they were asked to solve the second problem (ANA or INC) in 20 minutes by asking questions to \MuDoC. 
After the problem solving sessions, participants answered survey questions directly comparing \TeDoC~and \MuDoC, and some questions about their demographic background.
Finally, we conducted a brief semi-structured post-study interview where we asked them questions based on their survey responses.  
We did not shuffle the order of working with \TeDoC~and \MuDoC~because we found, during pilot studies, that participants who first interacted with \MuDoC~developed a bias as they attempted to emulate \MuDoC's feature such as ELI10 in \TeDoC~through prompting.
We documented detailed observations, participants' comments and questions, and verbatim quotes during the study and analyzed based on themes described in the next section.
More details on the surveys and problem-solving tasks are provided in the supplementary material\footref{footnote:samples-and-documentation}.

\section{Data Analysis and Results}\label{sec:results}

\subsection{Demographic and Background Information}
All participants were 18 to 35 years old.
12 had completed their bachelors degree while the other 18 also had a masters/professional degree.
Based on their reported experience, 6 participants were AI researchers, 12 were AI developers, and the remaining 12 were AI Users.

\subsection{Conversations with AI Systems}\label{subsec:conversations}
The participants in the study asked M=5.17, SD=2.40 questions to \MuDoC~ and M=5.86, SD=2.68 questions to \TeDoC, but a paired t-test revealed that this difference was not significant (t=1.481, p=0.149).
They spent an average of 19 minutes with both systems, which is close to 20 minutes time limit.

Regarding the difficulty of \ANA~and \INC~on a scale of 1 to 10, participants rated as follows: 
\ANA~(M=6.67, SD=1.19) was considered far more difficult (t=4.668, p<0.001) than \INC~(M=4.83, SD=1.53).
This likely explains a higher number of questions asked (t=2.469, p=0.019) for solving \ANA~problem (M=6.07, SD=2.48) compared to \INC~(M=4.97, SD=2.55)
and more time spent (t=3.737, p<0.001) on solving \ANA~problem (M=20.47, SD=2.7) compared to \INC~(M=18.50, SD=2.7).

The average response length excluding images for both systems was about 1750 characters (SD=634).
For \MuDoC, 58\% responses contained 1-5 images from the document
(M=2.02, SD=1.18). 
The average time-to-first-token for \TeDoC~and \MuDoC~was 4.1 and 6.9 seconds respectively.

\subsection{Survey Responses and Qualitative Feedback}\label{subsec:survey-responses}

\begin{figure}[t]
    \centering
    \includegraphics[width=0.85\textwidth]{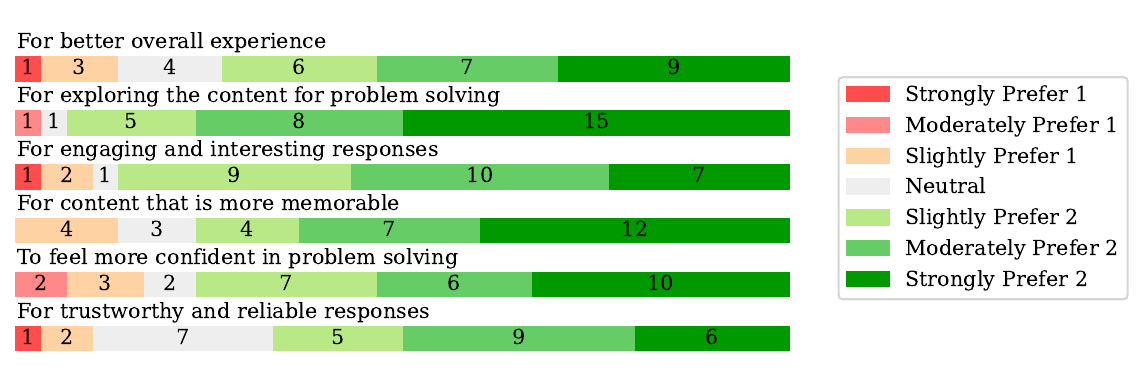}
    \caption{Participants preferences comparing text-only \TeDoC~(1) and multimodal \MuDoC~(2) system for learning experience. Numbers indicate participant count.} \label{fig:likert_comparison}
\end{figure}

Survey responses to statements comparing \TeDoC~and \MuDoC~on a 7-point Likert scale are summarized in Fig. \ref{fig:likert_comparison}.
Overall, 22 participants preferred \MuDoC~while only 4 preferred \TeDoC.
Out of 30, 28 preferred \MuDoC~\textit{for exploring the content for problem-solving}. 
Many explained that {navigation using text and images} allowed them to check responses for hallucinations, refer to the textbook frequently, examine surrounding text and figures, get more explanations from the book, and to create more informed follow-up questions.
One participant stated, 
\textit{``[\MuDoC~elements] are interactive-you're not just staring at content; you're navigating through it. It makes me want to spend more time with [\MuDoC].''} providing evidence that \MuDoC helped them engage with the content more actively compared to \TeDoC.

\textit{Regarding interest and engagement}, 26 participants preferred \MuDoC~while only 3 preferred \TeDoC.
Many characterized images in \MuDoC~responses as helpful in learning and effective in capturing attention.
Several participants also mentioned that \MuDoC~responses were more detailed and contained helpful examples
(\textit{``I liked the format of the answers, with the definitions and diagrams and bolded text—this is also how I take notes!''}).
Some participants preferred reading from the book and found navigation to be helpful and engaging. 
Those who preferred \TeDoC~found themselves more focused and engaged because of its minimalism and similarity to existing conversational AI interfaces.

Out of 30 participants, 23 perceived \MuDoC~responses to \textit{be more memorable,} while 4 favored \TeDoC.
Many believed that visuals in \MuDoC~responses made the content more memorable compared to text-only responses.
Several participants appreciated the interactive nature of the interface, noting that it encouraged deeper engagement with the content
(\textit{``[\MuDoC] made me think more about the answer and check if it was true. If there was a text or diagram that wasn't a part of the answer, it made me wonder why it wasn’t included. And when hallucinations are there, it lowers the barrier to verifying the text and being more cautious.''}).
Participants in favor of \TeDoC~believed that \MuDoC~ responses were less concise, which was essential for them because of time constraints.  

23 participants \textit{felt more confident about problem-solving} with \MuDoC~due to the
presence of visuals, 
ability to navigate to relevant content,
preference for reading from the book,
response credibility,
and ease of understanding due to the ELI10 feature.
However, participants who felt more confident with \TeDoC~expressed concerns about information overload when using \MuDoC.

Finally, 20 out of 30 participants favored \MuDoC~\textit{for trustworthiness and reliability of responses}~while only 3 preferred \TeDoC.
The ability to navigate to the source of images and texts using \MuDoC~was cited as the top reason for its higher credibility.
However, some participants raised concerns about \MuDoC's errors in navigation using text, and long responses with multiple images
(\textit{``Trustworthiness comes from concise and direct answers, [\MuDoC] seemed to be covering some bases and just gave too much information''}).

\textbf{Impact of Problem Order:}
To understand the impact of order of problems on user experience, we compared the response distribution using Mann-Whitney U test.
When participants solved the easier problem (INC) using \MuDoC, 
it led to favorable ratings for \MuDoC~in terms of being engaging (U=66.0, p=0.047), memorable (U=65.0, p=0.042), and perceived confidence (U=50.5, p=0.009).
While the participants who solved ANA using \MuDoC~(n=15) favored it on average, their responses were \textit{not} significantly better than neutral for overall experience (T=22.0, p=0.095), confidence (T=33.5, p=0.222), and trustworthiness (T=17.0, p=0.076) based on one-sample Wilcoxon test.

\begin{figure}[t]
    \centering
    \includegraphics[width=0.85\textwidth]{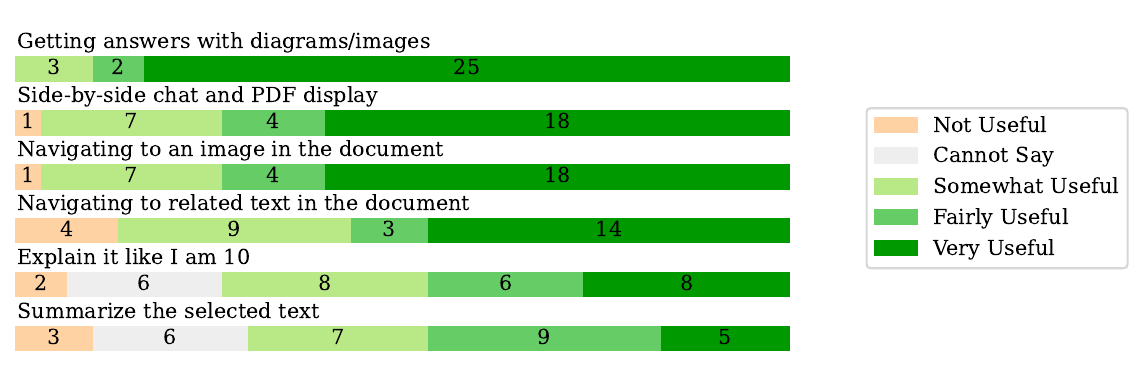}
    \caption{Perceived `usefulness' of different \MuDoC~features sorted by aggregated preferences. Numbers indicate participant count.} \label{fig:likert_features}
\end{figure}

\textbf{Feedback on UI}:
We asked participants to rate the perceived usefulness of six UI features of \MuDoC~on a 5-point Likert scale. 
The results are presented in Fig. \ref{fig:likert_features}. 
(1) All 30 participants found \MuDoC~\textit{responses with images} to be useful. 
Most believed that visuals were more effective in explaining concepts than text only 
(\textit{``Visual examples help a lot. For text, you have to mentally map it or write it while reading.''}).
One participant pointed out that images could not be hallucinated, as they were clipped from the document, highlighting their appreciation for grounding \MuDoC's responses in course documents.
However, some found multiple images to be overwhelming and the resulting responses to be too long. 
(\textit{``I felt as though [\MuDoC] provided more images than were necessary''}).
(2) 29 participants preferred the convenience of \textit{side-by-side chat and PDF display}, mainly because it enabled verification and increased trust.
Many explained that it helped them trust the AI response when they could confirm it by reading from the book.
(3, 4) 29 and 26 participants identified \textit{navigation using images and text} respectively as useful
due to the convenience in referring to the textbook, finding relevant content, and providing credibility to responses.
However, some expressed their frustration with navigating using text,
as few paragraphs in the AI responses were mapped to headings instead of text or to snippets that were not related.
(5, 6) 22 and 21 participants found ELI10 and Summarize features to be useful, respectively.
Many found them to be engaging and assistive for reading,
while others could not find enough time to utilize them.

\textbf{Other feedback:}
Some participants said that familiarity with text-only interfaces made it easier to interact with the \TeDoC. 
(\textit{``[\TeDoC] is more familiar, so I could use it instantly, but [\MuDoC] has more features that are useful which make it a more trustworthy system''}).
Long response lengths of both AI systems was another common concern
(\textit{``AI only comes in handy when you have [less] time, so it needs to provide short responses.''})

\subsection{Student Performance}
The first author and a teaching assistant (TA) of the Knowledge-based AI course created an initial rubric and 
independently evaluated $\approx$ 25\% of all solutions. 
Following this, the initial rubric was discussed again and adapted to create a robust final rubric.  
After independently evaluating all the solutions again, and scoring them on a scale of 0-12, all the disagreements were discussed to decide final scores.
We trained a linear mixed-effects model (LMM) using Python \texttt{statsmodels} package with participant scores as dependent variable, and system (\TeDoC~or \MuDoC) and problem (ANA or INC) as fixed effects.
Each group in LMM corresponds to a participant and has a random intercept.
Based on this model, we observe that the system did not impact the scores, \MuDoC~had a small but insignificant negative impact on student scores ($\hat{\beta}$=-0.350, SE=0.542, p=0.518).
On the other hand, \ANA~problem was more difficult and lead to a significant decrease in score ($\hat{\beta}$=-1.150, SE=0.542, p=0.034) compared to INC which is in agreement with average difficulty ratings discussed in Section \ref{subsec:conversations}.

\section{Discussion and Future Work}\label{sec:discussion}

In this section, we discuss our findings by grounding them in learning and cognitive science theories including MLT and CLT discussed in Section \ref{sec:related-work}. 
We also consider the implications for multimodal AI in education and accordingly provide directions for future research.    

\textbf{Multimedia Effect:}
We found that visuals in \MuDoC~responses were unanimously considered useful by the participants as visuals enhanced their learning experience and captured their attention. 
The \textit{multimedia effect} in MLT suggests that integration of verbal and visual channels by learning with text and images can lead to higher cognitive retention in students \cite{mayer_multimedia_2002}.
This was also concurred by participants as they perceived multimodal responses to be more memorable compared to text-only responses. 
Many visuals exemplified concepts and algorithms which positively influenced perception of \MuDoC~responses as learners appreciated examples along with abstract descriptions. 
This aligns with \textit{worked example effect} in CLT which suggests that students learn better when they are taught with worked examples instead of unguided problem solving \cite{sweller_cognitive_2011}. 
While higher cognitive retention owed to multimedia learning and worked example effect have been established before, more research is required to realize them in multimodal human-AI interactions because of different conditions such as learner goals, time constraints, passive versus active role, and levels of personalization.

\textbf{Reliability and Credibility:}
\MuDoC~responses were considered as more credible 
because participants could examine the source of responses on a granular level.
The use of document-grounded images also increased trust because \MuDoC~images could not be hallucinated, and their source was easily accessible.
The document next to the chat also supported this with increased convenience without a need for additional steps like switching tabs.
In information seeking, reliability (consistent high quality responses) and credibility (believability of information or its source) are among the most important predictors of trust \cite{thielsch_trust_2018}. 
This explains the difference between perceived trust in \MuDoC~and \TeDoC~as the latter does not provide credible evidence for its responses.
However, as indicated by some participants, \TeDoC~could be perceived as more reliable for its consistent response length, which was not inflated by images.
Further, the presence of sources can improve trust, but quality of source attribution is also important if users examine these sources \cite{ding_citations_2025}.
Consequently, reliability of \MuDoC-like systems can be improved by providing concise responses that focus on student queries rather than coverage of retrieved content, and through reliable source attribution. 
Future work should explore methods to obtain granular feedback, 
for instance, by asking learners to highlight useful content in responses, to communicate their priorities and understanding.
Such information can be capitalized to generate succinct personalized responses.

\textbf{Learning Outcomes:}
The text in \MuDoC~responses had the same average length as \TeDoC, but additional information consisted of about two images on average and source mapping to the document.
The \textit{split-attention effect} in CLT \cite{sweller_cognitive_2011} suggests that integrating information from multiple sources increases cognitive load which can explain the information overload experienced by some participants.
Further, \textit{environmental organizing and linking principle} in CLT suggests that long-term memory, such as familiarity with an AI system, can reduce load on working memory and improve learning \cite{sweller_cognitive_2011}.
Some participants indeed acknowledged the potential impact of unfamiliarity with \MuDoC~on their experience. 
While individual effect of these factors can be small, the total cognitive load may exceed working memory capacity 
which can explain the small (non-significant) negative trend in scores due to \MuDoC.

\textbf{Critical Thinking:}
With navigation using text and images, participants were encouraged to explore the document more frequently. 
The exploration empowered them to check responses for hallucinations and to delve deeper by examining related content.
This suggests that the navigation features in \MuDoC~helped participants to actively question and evaluate AI-generated content, rather than passively accepting its responses, suggesting that that \MuDoC~ can promote critical thinking.
When considered as a skill, critical thinking requires deliberate practice \cite{gelder_teaching_2005} and future work could examine whether AI systems that enable such forms of critical evaluation can positively impact learner behavior in the long term.
We believe that this is an important step towards mitigating over-reliance on generative AI tools for effective learning.

\textbf{Cognitive Processing:}
Interestingly, some participants noted that verifying the AI responses can be helpful for longer retention of the concepts they learned.
This could be explained by Craik and Lockhart's \textit{levels of processing} framework of memory \cite{craik_levels_1972}, according to which learning activities that require higher cognitive processing are more effective for long-term memory compared to mere repetition. 
Existing AI systems that provide sources require many additional steps to verify their correctness, due to which most users do not frequently engage in source verification. \cite{ding_citations_2025}. In contrast, \MuDoC~affords greater interactivity due to the ease of source verification. We hypothesize that this positively contributed to higher cognitive processing, and, therefore, higher perceived retention.

\textbf{Load Regulation:}
In contrast to positive impact on cognitive retention, CLT suggests that processing multiple novel elements can lead to poor learning if cognitive load exceeds working memory capacity \cite{sweller_cognitive_2011}.
This could explain why some participants felt overwhelmed by \MuDoC's responses.
It is therefore crucial to design AI tools to strike a balance between cognitive engagement and cognitive load in multimedia learning.
AI tools should allow learners to self-regulate their cognitive load.
Future work should explore how to afford this autonomy through options to disable certain functionalities or to hide interface elements as needed.

\textbf{Summary:}
\MuDoC~provides engaging responses with interleaved text and visuals that have the potential to improve learning outcomes such as cognitive retention, in agreement with MLT. 
Trustworthiness in AI systems for information-seeking tasks depends on its ability to provide consistent high quality responses and to support source verification. 
For \MuDoC, document-grounded visuals and navigation using text and images lead to increased credibility.
However, more work is needed to generate concise personalized AI responses based on learner models and to improve granular source attribution.
The active process of verification, when convenient to perform, leads to higher engagement and may positively influence critical thinking and cognitive retention. 
To further leverage such interactive systems for student learning, there is a need to design systems that allow self-regulation of cognitive load by providing students greater autonomy.

\textbf{Limitations:}
In our study, participants learned novel concepts from an AI system instead of direct instruction from instructors. 
More experimentation is needed to understand the effect of prior instruction on problem-solving using AI systems.
\MuDoC's novel features may have required a longer warm-up period to reach a familiarity level at par with text-only interfaces.  
Future studies can also benefit from providing more time for learning and problem-solving in randomized controlled trials with a larger participant pool.  
Finally, all participants in our study were graduate students at an acclaimed university which is a non-representative group in terms of age and academic background.

\section{Conclusion}
We presented a multimodal document-grounded conversational AI system that employs text and visuals from documents to support multimedia learning through conversations.
We conducted a study to compare our system with a text-only system in a realistic setting where participants solved problems while learning by asking questions to an AI system.
Our results show that multimodal responses lead to higher engagement compared to text-only responses and the interface promoted trustworthiness as it allowed verification of text and image content through seamless navigation.
While we did not observe a significant change in performance on problem solving tasks, 
the qualitative feedback suggests that (1) illustrative visuals and cognitive engagement through verification can improve cognitive retention and critical thinking skills, and (2) document-grounded visuals and verifiability increase credibility and trust in AI systems.
We discussed our findings in light of theories of learning and cognition to understand implications for multimodal AI for education. 
We believe that multimodal conversational AI will play an important role in improving learning experience and our work is an important step forward in highlighting its benefits and current challenges.

\section*{Acknowledgments} 

We are grateful for the support provided by National Science Foundation under Grant No. 2247790 and Grant No. 2112532. Any opinions, findings, and conclusions or recommendations expressed in this material are those of the author(s) and do not necessarily reflect the views of the National Science Foundation.

We also wish to thank John Kos, a teaching assistant for the Knowledge-based AI course, for helping with evaluation of participants' answers in the user study.
We also thank Sandeep Kakar and Lingqing Wang for their feedback on the user study design.

\bibliographystyle{splncs04}
\bibliography{references}

\end{document}